\documentclass[twocolumn,showpacs,preprintnumbers,amsmath,amssymb,superscriptaddress,showkeys,prd]{revtex4}

\usepackage{graphicx}
\usepackage{dcolumn}
\usepackage{bm}
\usepackage{amsmath,amssymb}

\begin{document}

\preprint{APS/123-QED}

\title{Constituent quark masses obtained from hadron masses with contributions of Fermi-Breit and Glozman-Riska hyperfine interactions}

\author{V. Borka Jovanovi\'{c}}
\email[Corresponding author:]{vborka@vinca.rs}
\affiliation{Laboratory of Physics (010), Vin\v{c}a Institute of
Nuclear Sciences, University of Belgrade, P.O. Box 522, 11001
Belgrade, Serbia}

\author{S. R. Ignjatovi\'{c}}
\affiliation{Department of Physics, Faculty of Science, University of Banja Luka, Mladena
Stojanovi\'{c}a 2, 78000 Banja Luka, Bosnia and Herzegovina}
\author{D. Borka}

\affiliation{Laboratory of Physics (010), Vin\v{c}a Institute of
Nuclear Sciences, University of Belgrade, P.O. Box 522, 11001
Belgrade, Serbia}

\author{P. Jovanovi\'{c}}
\affiliation{Astronomical Observatory, Volgina 7, 11060 Belgrade,
Serbia}

\date{December 3, 2010}

\begin{abstract}
We use the color-spin and flavor-spin interaction Hamiltonians with
SU(3) flavor symmetry breaking to obtain meson and baryon mass
formulas. Adjusting these masses with experimental masses we
determine the constituent quark masses. We discuss the constituent
quark masses obtained from meson and baryon mass fits. The results
for constituent quark masses are very similar in case of two
different phenomenological models: Fermi-Breit and Glozman-Riska
hyperfine interactions.
\end{abstract}



\maketitle

\section{Introduction}

The quark model predicts that the masses of the mesons and baryons
are given by the sum of the constituent masses of the quarks and the
hyperfine splitting \cite{lave97,gasi81,povh95}. As described in
\cite{lave97}, the term with hyperfine splitting emanates from a
spin-spin interaction produced by one gluon exchange, which also
depends on the masses of the quarks (for mesons this term contains a
mass sum of a quark pair and for baryons it contains a sum over all
the possible quark pairs).

In Refs. \cite{isgu79isgu79} the baryons were studied in a quark
model with color hyperfine interactions. Some parameters of meson
and baryon spectroscopy, including some predictions of the quark
masses, can be found in \cite{godf85}. In Ref. \cite{karl03}, from
the meson and baryon relations, the authors fitted both the mass
differences and mass ratios with a single set of quark masses. In
\cite{scad06} estimations of masses were made via magnetic moments
of baryons and via the vector meson masses. In Refs.
\cite{ronc95a,ronc95b} it was shown that in the constituent quark
model, the Feynman-Hellmann theorem and semiempirical mass formulas
can be applied to give useful information about the masses of mesons
and baryons.

In this paper we use a schematic study (two quark interaction) to
calculate meson and baryon theoretical masses from which the
constituent quark masses can be obtained. This paper is organized in
the following way: in Section II we give complete method how to
calculate theoretical masses, in Section III we explain fitting
procedure and our main results, discussion and conclusions are given
in Section IV.

\section{Contribution of the hyperfine interactions}

Using the colored version of the Fermi-Breit hyperfine interaction
(FB HFI) \cite{deru75,libe77,luch91,silv92} and the Glozman-Riska
hyperfine interaction (GR HFI) \cite{gloz96,bork07,bork08}, we
present estimations of the theoretical meson and baryon masses.
Strong FB HFI Hamiltonian \cite{silv92} has the following form:
\begingroup
\setlength{\abovedisplayskip}{0pt}
\setlength{\belowdisplayskip}{0pt}
\begin{equation}
H_{\rm FB} = C\sum\limits_{i > j} {\frac{\vec \sigma_i \vec
\sigma_j}{m_i m_j}\left( {\lambda_i^C \lambda_j^C} \right)},
\label{equ01}
\end{equation}
\endgroup
\noindent and it has explicit color and spin exchange dependence and
implicit (by way of quark masses) flavor dependence. Here $\sigma_i$
are the Pauli spin matrices, $\lambda^C_i$ are the color Gell-Mann
matrices and $C$ is a constant. FB contribution to hadron masses is
given by: $ m_{\nu,\rm FB} = \left\langle \nu \right|\left\langle
\chi \right|H_{\rm FB} \left| \chi \right\rangle \left| \nu
\right\rangle$, where $\chi$ denotes the spin wave function and
$\nu$ -- the flavor wave function. Strong GR Hamiltonian
\cite{gloz96} is of the form:
\begingroup
\setlength{\abovedisplayskip}{0pt}
\setlength{\belowdisplayskip}{0pt}
\begin{equation}
\begin{array}{r}
H_\mathrm{GR} = - \mathrm{C_\chi} \sum\limits_{i < j} {\left( -1 \right)^{\mathop \alpha_{ij}}
\left( {\lambda_i^F \lambda_j^F } \right)\left( \frac{\vec \sigma _i \vec \sigma_j}{m_i m_j}\right)},\\
\left( -1 \right)^{\mathop \alpha_{ij}} = \left\{
{\begin{array}{*{20}c} {\begin{array} {*{20}c}{ -1,} & {q\bar q} \\
\end{array}} \\
{\begin{array}{*{20}c} { +1,} & {qq \ {\rm or}\ \bar q\bar q} \\
\end{array}} \\
\end{array}} \right\}
,\end{array}
\label{equ02}
\end{equation}
\endgroup
\noindent where $\lambda_{i}^{F}$ are Gell-Mann matrices for flavor
SU(3), $\sigma_{i}$ are the Pauli spin matrices and C$_{\chi}$ is a
constant. We employ this schematic flavor-spin interaction between
quarks and antiquarks which leads to Glozman-Riska HFI contribution
to hadron masses: $m_\mathrm{\nu,GR} = \left\langle \nu
\right|\left\langle \chi \right|H_{\rm GR} \left| \chi \right\rangle
\left| \nu \right\rangle$, where m$_{i}$ are the constituent quark
effective masses: $m_\mathrm{u}$ = $m_\mathrm{d}$  $\neq$
$m_\mathrm{s}$ and $\nu$ - flavor wave functions.

For mesons, there are two flavor SU(3) multiplets according to
product: $3 \otimes \bar 3 = 1 + 8$, i.e.~one singlet and one octet.
For baryons, according to: $3 \otimes 3 \otimes 3 = 1 + 8_{MS} +
8_{MA} + 10$, there is one singlet, two octets (one mixed symmetric
(MS) and one mixed antisymmetric (MA)) and one decuplet. Young
diagrams for these SU(3)$_\mathrm{F}$ multiplets, as well as the
weight diagrams, can be found in textbooks (see
e.g.~\cite{sche05,halz84}). Regarding baryons, in case of octet
members (as they have the MS and MA part of the flavor wave
function), the contribution of HFI is calculated as sum of the
symmetric and antisymmetric part, so we start from the following
expression: $\left| {\nu \uparrow} \right\rangle = \frac{1}{\sqrt
2}\left| {\nu_{MS} + \nu_{MA}} \right\rangle$.

\section{Fitting procedure}

For our calculations we used multidimensional least-square fit of
masses, using subroutine "lfit" from Numerical Recipes in FORTRAN
\cite{pres92}, modified according to the instructions in the last
paragraph of \S 15.4 of \cite{pres92}. We fitted simultaneously all
parameters: $m_u$, $m_s$, $m_c$ and the constant $C$ in such a way
that we minimized $\chi^2$ between the measured and theoretical
masses. Equations for meson and baryon theoretical masses (see \S
IV) are at first linearized by expansion in Taylor's series up to
the first order, and in that way we obtained the corresponding
system of linear equations for differences between experimental and
theoretical masses in which the unknown variables are corrections of
the parameters. These corrections, as well as their corresponding
uncertainties, are determined using linear least square method, and
then the parameters are accounted for the values of these obtained
corrections. With these new (corrected) parameters the previous
procedure is repeated until the fit converges, that is while the
$\chi^2$ value between the experimental and theoretical masses
decreases. In that way, after several iterations, we obtain final
values for the parameters and their uncertainties. The uncertainties
are estimated during fitting procedure as square roots of the
corresponding diagonal elements of covariance matrix, according to
eq. (15.4.15) of Ref. \cite{pres92}.

For every analyzed system of equations a fast convergence is
obtained, even in the case when starting values of parameters differ
very much from their final values, which favors the correctness of
our theoretical model as well as our fitting method.

\section{Results and Discussion}

We calculated FB and GR contributions to meson and baryon masses.
The mass formulas of mesons and baryons with FB HFI included are
given in eqs.~(\ref{equ03}) - (\ref{equ04}), while their mass
formulas with GR HFI included can be found in \cite{bork07}
(eqs.~(6)-(10) for mesons and eqs.~(11)-(13) for baryons).

Here we study the hadron masses using FB and GR HFI in schematic
approximation (two-particle interaction). The masses of constituent
quarks $m_u$ (= $m_d$), $m_s$, $m_c$ and the constants $C^m$ and
$C^b$ are calculated from least-square fit of the theoretical
equations for meson and baryon masses with HFI included. The
corresponding experimental masses are taken from "Particle Data
Group" site: \url{http://pdg.lbl.gov} \cite{PDG10}.

We give the equations for meson masses with FB interaction included
for the following mesons: light pseudoscalar mesons $\pi$, $K$,
light vector mesons $\rho$, $K^*$, $\omega$, $\varphi$, charmed
mesons $D$, $D^*$, strange charmed mesons $D_S$, $D_S^*$ and double
charmed mesons $\eta_\mathrm{c}$, J/$\psi$ (eqs.~(\ref{equ03})). In
these equations, the constant for mesons is denoted by $C^m$. We did
not calculate $\eta$ and $\eta'$ contribution because of their
mixing and they cannot be described within such a model. Because of
that mixing (due to the same quantum numbers), their flavor wave
functions are given only in a first approximation \cite{jaff77a} and
therefore calculations are not sufficiently precise. The mixing of
the states also changes the properties and shifts masses from the
theoretical predictions.

\begingroup
\setlength{\abovedisplayskip}{0pt}
\setlength{\belowdisplayskip}{0pt}
\begin{equation}
\begin{array}{l}
m_\pi = 2m_u - \dfrac{3C^{m}}{m_u^2};
m_K = m_u + m_s - \dfrac{3C^{m}}{m_u m_s} \\
m_\rho = 2m_u + \dfrac{C^{m}}{m_u^2};
m_{K*} = m_u + m_s + \dfrac{C^{m}}{m_u m_s} \\
m_\omega = 2m_u + \dfrac{C^{m}}{m_u^2};
m_\varphi = 2m_s + \dfrac{C^{m}}{m_s^2} \\
m_D = m_u + m_c - \dfrac{3C^{m}}{m_u m_c};
m_{D*} = m_u + m_c + \dfrac{C^{m}}{m_u m_c} \\
m_{Ds} = m_s + m_c - \dfrac{3C^{m}}{m_s m_c};
m_{Ds*} = m_s + m_c + \dfrac{C^{m}}{m_s m_c} \\
m_{\eta c} = 2m_c - \dfrac{3C^{m}}{m_c^2};
m_{J/\psi} = 2m_c + \dfrac{C^{m}}{m_c^2}.
\end{array}
\label{equ03}
\end{equation}
\endgroup

\begin{table*}
\centering
\caption{Constituent quark masses obtained from the meson
fits. For each combination of mesons, the results are obtained from
their masses with FB HFI included (upper rows) and with GR HFI
(lower rows). Note that constant $C^m$ differs for the two HFIs.}
\begin{ruledtabular}
\begin{tabular}{lcccccc}
\noalign{\smallskip}
Fit & Mesons & & Quark masses (MeV) & & $C^m$ & $\chi^2$ \\
No. & & $m_{u}$ = $m_{d}$ & $m_{s}$ & $m_{c}$ & ($\times 10^7 \mathrm{MeV}^3$) & ($\times 10^6$) \\
\noalign{\smallskip}
\hline
\noalign{\smallskip}
1 & $\pi$, $K$, $\rho$, $K^*$, $\omega$, $\varphi$, $D$, & 318.12 $\pm$ 0.11 & 477.29 $\pm$ 0.21 & 1591.42 $\pm$ 0.21 & 1.5977 $\pm$ 0.0017 & 0.095336 \\
\cline{3-7}
& $D^*$, $D_S$, $D_S^*$, $\eta_\mathrm{c}$, J/$\psi$ & 304.55 $\pm$ 0.11 & 534.33 $\pm$ 0.19 & 1576.94 $\pm$ 0.21 & 2.1331 $\pm$ 0.2346 & 0.304832 \\
\noalign{\smallskip}
\hline
\noalign{\smallskip}
2 & $\pi$, $K$, $\rho$, $K^*$, $\omega$, $\varphi$ & 308.05 $\pm$ 0.12 & 484.91 $\pm$ 0.24 & - & 1.5043 $\pm$ 0.0017 & 0.000411 \\
\cline{3-7}
&  & 292.35 $\pm$ 0.12 & 552.80 $\pm$ 0.22 & - & 2.0284 $\pm$ 0.0023 & 0.143289 \\
\noalign{\smallskip}
\hline
\noalign{\smallskip}
3 & $D$, $D^*$, $D_S$, $D_S^*$, & 454.75 $\pm$ 0.47 & 547.48 $\pm$ 0.49 & 1524.36 $\pm$ 0.35 & 2.6936 $\pm$ 0.0095 & 0.003475 \\
\cline{3-7}
& $\eta_\mathrm{c}$, J/$\psi$ & 453.68 $\pm$ 0.44 & 546.82 $\pm$ 0.49 & 1524.94 $\pm$ 0.35 & 4.0280 $\pm$ 0.0138 & 0.003480 \\
\noalign{\smallskip}
\end{tabular}
\end{ruledtabular}
\label{tab01}
\end{table*}

\begin{table*}
\centering
\caption{The same as Table I, but for baryon fits. Note
that for heavy baryons fit did not converge and constitutive quark
masses could not be obtained.}
\begin{ruledtabular}
\begin{tabular}{lcccccc}
\noalign{\smallskip}
Fit & Baryons & & Quark masses (MeV) & & $C^b$ & $\chi^2$ \\
No. & & $m_{u}$ = $m_{d}$ & $m_{s}$ & $m_{c}$ & ($\times 10^7 \mathrm{MeV}^3$) & ($\times 10^6$) \\
\noalign{\smallskip}
\hline
\noalign{\smallskip}
1 & $N$, $\Sigma$, $\Xi$, $\Lambda$, $\Delta$, $\Sigma^{*}$, $\Xi^{*}$, & 381.53 $\pm$ 0.09 & 537.26 $\pm$ 0.18 & 1368.15 $\pm$ 0.27 & 1.3454 $\pm$ 0.0021 & 0.153491 \\
\cline{3-7}
& $\Omega$, $\Sigma_\mathrm{c}$, $\Xi_\mathrm{cc}$, $\Lambda_\mathrm{c}$, $\Sigma_\mathrm{c}^{*}$, $\Omega_\mathrm{c}$ & 541.04 $\pm$ 0.18 & 655.00 $\pm$ 0.22 & 1358.62 $\pm$ 0.25 & 2.4878 $\pm$ 0.0047 & 0.125597 \\
\noalign{\smallskip}
\hline
\noalign{\smallskip}
2 & $N$, $\Sigma$, $\Xi$, $\Lambda$, $\Delta$, $\Sigma^{*}$, $\Xi^{*}$, & 362.52 $\pm$ 0.10 & 538.68 $\pm$ 0.20 & - & 1.2789 $\pm$ 0.0020 & 0.001165 \\
\cline{3-7}
& $\Omega$ & 500.42 $\pm$ 0.20 & 621.20 $\pm$ 0.25 & - & 1.6951 $\pm$ 0.0043 & 0.002719 \\
\noalign{\smallskip}
\end{tabular}
\end{ruledtabular}
\label{tab02}
\end{table*}

\begin{table*}[ht!]
\centering
\caption{Constituent quark masses obtained from hadron
fits when hadrons with two $c$-quarks are excluded. The upper rows
correspond to FB HFI, and the lower rows to GR HFI.}
\begin{ruledtabular}
\begin{tabular}{lcccccc}
\noalign{\smallskip}
 & Hadrons & & Quark masses (MeV) & & $C$ & $\chi^2$ \\
 & & $m_{u}$ = $m_{d}$ & $m_{s}$ & $m_{c}$ & ($\times 10^7 \mathrm{MeV}^3$) & ($\times 10^6$) \\
\noalign{\smallskip}
\hline
\noalign{\smallskip}
Mesons & $\pi$, $K$, $\rho$, $K^*$, $\omega$, $\varphi$, $D$, & 314.75 $\pm$ 0.12 & 466.80 $\pm$ 0.21 & 1627.31 $\pm$ 0.25 & 1.5546 $\pm$ 0.0017 & 0.033123 \\
\cline{3-7}
& $D^*$, $D_S$, $D_S^*$, $\Lambda_\mathrm{c}$, $\Sigma_\mathrm{c}^{*}$, $\Omega_\mathrm{c}$ & 300.88 $\pm$ 0.11 & 526.65 $\pm$ 0.19 & 1605.62 $\pm$ 0.26 & 2.0660 $\pm$ 0.0023 & 0.263221 \\
\noalign{\smallskip}
\hline
\noalign{\smallskip}
Baryons & $N$, $\Sigma$, $\Xi$, $\Lambda$, $\Delta$, $\Sigma^{*}$, $\Xi^{*}$, & 365.69 $\pm$ 0.09 & 530.08 $\pm$ 0.18 & 1700.17 $\pm$ 0.37 & 1.2513 $\pm$ 0.0019 & 0.015451 \\
\cline{3-7}
& $\Omega$, $\Sigma_\mathrm{c}$, $\Lambda_\mathrm{c}$, $\Sigma_\mathrm{c}^{*}$, $\Omega_\mathrm{c}$ & 506.26 $\pm$ 0.19 & 623.40 $\pm$ 0.22 & 1629.69 $\pm$ 0.35 & 1.8163 $\pm$ 0.0041 & 0.032378 \\
\noalign{\smallskip}
\end{tabular}
\end{ruledtabular}
\label{tab03}
\end{table*}

\begin{table*}[ht!]
\centering
\caption{Predicted masses (in MeV) of constituent quarks
in different phenomenological models.}
\begin{ruledtabular}
\begin{tabular}{lccccccc}
\noalign{\smallskip}
Mass & Ref. \cite{godf85} & Ref. \cite{ronc95a} & Ref. \cite{lave97} & Ref. \cite{bura98} & Ref. \cite{karl03} & Ref. \cite{scad06} & Ref. \cite{bork07} \\
\noalign{\smallskip}
\hline
\noalign{\smallskip}
$m_u$ = $m_d$ & 220 & 300 & 310 & 290 & 360 & 337.5 & 311 \\
\noalign{\smallskip}
\hline
\noalign{\smallskip}
$m_s$ & 419 & 475 & 483 & 460 & 540 & 486 & 487 \\
\noalign{\smallskip}
\hline
\noalign{\smallskip}
$m_c$ & 1628 & 1640 & - & 1650 & 1710 & 1550 & 1592 \\
\noalign{\smallskip}
\end{tabular}
\end{ruledtabular}
\label{tab04}
\end{table*}

We also present the theoretical mass equations with FB HFI included
for the following baryons: light baryon octet $N$, $\Sigma$, $\Xi$,
$\Lambda$, light baryon decuplet $\Delta$, $\Sigma^{*}$, $\Xi^{*}$,
$\Omega$ and heavy baryons $\Sigma_\mathrm{c}$, $\Xi_\mathrm{cc}$,
$\Lambda_\mathrm{c}$, $\Sigma_\mathrm{c}^{*}$, $\Omega_\mathrm{c}$
(eqs.~(\ref{equ04})). Constant for baryons is denoted by $C^b$.

\begingroup
\setlength{\abovedisplayskip}{0pt}
\setlength{\belowdisplayskip}{0pt}
\begin{equation}
\begin{array}{l}
m_N = 3m_u - 3C^b \dfrac{1}{2m_u ^2} \\
m_\Sigma = 2m_u + m_s + 2C^b \dfrac{1}{m_u^2}\left( \dfrac{1}{4} - \dfrac{m_u}{m_s} \right) \\
m_\Xi = m_u + 2m_s + 2C^b \dfrac{1}{m_s^2}\left( \dfrac{1}{4} - \dfrac{m_s}{m_u} \right) \\
m_\Lambda = 2m_u + m_s - 3C^b \dfrac{1}{2m_u^2} \\
m_\Delta = 3m_u + 3C^b \dfrac{1}{2m_u^2} \\
m_{\Sigma *} = 2m_u + m_s + C^b \dfrac{1}{m_u^2} \left( \dfrac{1}{2} + \dfrac{m_u}{m_s} \right) \\
m_{\Xi *} = m_u + 2m_s + C^b \dfrac{1}{m_s^2} \left( \dfrac{1}{2} + \dfrac{m_s}{m_u} \right) \\
m_\Omega = 3m_s + 3C^b \dfrac{1}{2m_s^2} \\
m_{\Sigma c} = 2m_u + m_c + 2C^b \dfrac{1}{m_u^2} \left( \dfrac{1}{4} - \dfrac{m_u}{m_c} \right) \\
m_{\Xi cc} = m_u + 2m_c + 2C^b \dfrac{1}{m_c^2} \left( \dfrac{1}{4} - \dfrac{m_c}{m_u} \right) \\
m_{\Lambda c} = 2m_u + m_c - 3C^b \dfrac{1}{2m_u^2} \\
m_{\Sigma *c} = 2m_u + m_c + C^b \dfrac{1}{m_u ^2} \left( \dfrac{1}{2} + \dfrac{m_u}{m_c} \right) \\
m_{\Omega c} = 2m_s  + m_c + C^b \dfrac{1}{m_s^2} \left( \dfrac{1}{2} + \dfrac{m_s}{m_c} \right).
\end{array}
\label{equ04}
\end{equation}
\endgroup

\noindent For each set of equations, the minimized $\chi^{2}$ values
for masses are calculated by the formula: $\chi^2 = \sum\nolimits_{i
= 1}^N {\frac{{{\left( {T_i - E_i} \right)}^2}}{\sigma_i^2}}$, where
$T_i$ is the model prediction for the hadron mass, $E_i$ the
experimental hadron mass and $\sigma_i$ the uncertainty of the mass.

The values of constituent quark masses and their uncertainties (in
MeV) are given in Tables \ref{tab01} and \ref{tab02}. It is
noticeable that the fits are satisfactory for both HFIs and in cases
of all mesons, light mesons and light baryons. They show that the
constituent mass is only slightly modified by the dynamics of
confinement. The constituent masses differ for the quarks in mesons
and those in baryons. This is not unexpected since they are confined
in quite different systems. Also, as noted in \cite{ronc95a} there
are no theoretical reasons why the masses determined from the baryon
fits should coincide exactly with those determined from the meson
fits because these quark masses are constitutive masses, i.e.
effective ones. Roncaglia et al. \cite{ronc95a,ronc95b} also showed
that constraints (inequalities) for mass differences are stronger
for baryons than for mesons, and therefore requirement for the same
set of quark masses for both baryons and mesons would result in
significantly poorer fits \cite{ronc95a}.

In the case of heavy mesons, the fit resulted in somewhat increased
values for $u$ and $s$ quarks, and in the case of heavy baryons fit
even did not converge. Therefore, our results suggest that both of
the analyzed HFIs can be used in a first approximation for modeling
the term with hyperfine splitting in the quark model only for light
mesons and light baryons. That is both HFIs well describe systems
which contain quarks from SU(3)$_\mathrm{F}$, i.e. light quarks:
$u$, $d$, $s$. In order to verify this conclusion, we also show the
results for the case when hadrons with two $c$-quarks are excluded
(see Table \ref{tab03}): meson fit without $\eta_\mathrm{c}$ and
J/$\psi$, and baryon fit without $\Xi_\mathrm{cc}$. From these two
fits we can calculate mass differences $m_s - m_u$ and $m_c - m_s$
which may be then compared with the corresponding inequalities given
by Roncaglia et al. \cite{ronc95a,ronc95b}, obtained from the
Feynman-Hellmann theorem. We can see from Table \ref{tab03} that in
this case the constituent quark masses obtained for FB HFI satisfy
all of the inequalities given in \cite{ronc95a,ronc95b}, contrary to
Tables \ref{tab01} and \ref{tab02}, where FB HFI breaks some of
these inequalities. Also, from Table \ref{tab03} one can see that GR
HFI violates one or more of the inequalities, but to a lesser extent
than in Tables \ref{tab01} and \ref{tab02}. Even more, in this case
the obtained constituent quark masses are more realistic than those
from Tables \ref{tab01} and \ref{tab02}. From each set of equations
(except for heavy baryons) we calculated masses of constituent
quarks and constants $C^m$ and $C^b$, and we got slightly different
values, but all of them are from expected mass ranges of constituent
quarks. As one can see from Table \ref{tab04}, our predictions for
constituent quark masses are in the range of masses obtained using
different phenomenological models. When comparing these two HFIs, it
is interesting to note that the values of obtained quark masses are
similar although one interaction is color-spin, and the other one is
flavor-spin. As it can be seen from Tables \ref{tab01} and
\ref{tab02}, two different HFIs result with masses of constituent
quarks which are more similar than those obtained by the same HFIs
applied to different groups of hadrons (e.g. light mesons, heavy
mesons, light baryons, ...). For example, results for $m_u$ (=
$m_d$) and $m_s$ for FB HFI applied on light and heavy mesons are
more different than results for $m_u$ (= $m_d$) and $m_s$ for FB and
GR HFIs applied on light mesons.

When comparing FB, the color-spin interaction, with GR HFI, which is
flavor-spin interaction, it is interesting that we obtained similar
results. We can conclude that both HFIs describe masses of light
mesons and baryons very well. It shows that both HFIs well describe
systems which contains quarks from SU(3)$_\mathrm{F}$. In case of
heavy baryons the fits for both HFIs did not converge, which might
indicate that quark model which includes these two HFIs is not
satisfactory approximation for heavy baryons. It might also indicate
that FB and GR HFI are not the complete effective two-quark
interactions and therefore they cannot be successfully applied to
quark systems out of SU(3) group. If we do not take into account
hadrons with two $c$-quarks the FB HFI becomes a good approximation,
even for hadrons having one heavy $c$-quark. We also show that the
constituent quark masses are very sensitive to the system in which
they are confined, and their values differ less or more in different
systems (for example heavy mesons and light mesons). We obtained
that least-square fit is more accurate for FB than for GR
interaction in case of light mesons, but in case of heavy mesons and
light baryons accuracy of the fit is similar.

\vspace{-0.5cm}

\begin{acknowledgments}
This research is supported by the Ministry of Science of the
Republic of Serbia through project No. 176003.
\end{acknowledgments}


\end{document}